\documentclass[page-classic]{epl2}
\usepackage{graphicx}
\usepackage{psfrag}

\title{Internal chaos in an open quantum system: \\
From Ericson to conductance fluctuations}

\author{S.~Sorathia\inst{1} \and F.M.~Izrailev\inst{1,2} \and
  G.L.~Celardo\inst{3} \and V.G.~Zelevinsky\inst{2} \and G.P.~Berman\inst{4}}

\institute{
  \inst{1} Instituto de F\'{\i}sica, Universidad Aut\'{o}noma de
Puebla - Apartado Postal J-48, Puebla, Pue., 72570, M\'{e}xico\\
  \inst{2} NSCL and Dept. of Physics and Astronomy, Michigan State
University - East Lansing, Michigan 48824-1321, USA \\
  \inst{3} Physics Department, Tulane University - New Orleans, LA 70118, USA \\
  \inst{4} Theoretical Division and CNLS, Los Alamos National
Laboratory - Los Alamos, New Mexico 87545, USA
}


\abstract{
The model of an open Fermi-system is used for studying
the interplay of intrinsic chaos and irreversible decay
into open continuum channels. Two versions of the model are
characterized by one-body chaos coming from disorder or by
many-body chaos due to the inter-particle interactions.
The continuum coupling is described by the effective
non-Hermitian Hamiltonian. Our main interest is in specific
correlations of cross sections for various channels in
dependence on the coupling strength and degree of internal
chaos. The results are generic and refer to common features
of various mesoscopic objects including conductance
fluctuations and resonance nuclear reactions.
}

\begin{document}

\maketitle

\section{Introduction}
The problem of quantum transport is generic for all realistic quantum
systems interacting with environment. A transmission of a signal through
a many-body quantum aggregate of interacting particles is essentially
the main instrument in studying such systems and using them for practical
communication purposes. Currently this is one of the crucial lines of
development of mesoscopic physics \cite{MK04} with broad applications
to quantum information, electronics and material science.

Historically, many ideas nowadays defining mesoscopic physics emerged
in nuclear theory starting with Bohr's concept of reactions proceeding
through compound nucleus. Low-energy neutron resonances in heavy nuclei
present a typical example of exceedingly complex quasistationary states
in an open many-body system which serve as intermediaries in reaction
processes. Later these states provided the statistical justification for
the ideas of quantum chaos based on random matrix theory \cite{brody81}.
The detailed reviews of progressing knowledge on quantum chaos in complex
atoms and nuclei can be found in \cite{grib94,big,guhr98,PW07}; general
features of mesoscopic systems of interacting fermions were stressed
in \cite{FI97}. The theoretical concepts of many-body quantum chaos are
convincingly supported by the large-scale diagonalization of Hamiltonian
matrices. Although in mesoscopic condensed matter systems one-body chaos
often plays the main role, the interaction effects are also important and
the analysis has many features parallel to the nuclear theory
\cite{A90,AGKL97}.

When the lifetime of quasistationary states is getting small and
the corresponding resonances overlap, the openness of the system becomes
a decisive factor. In nuclear reactions this regime is called Ericson
fluctuations \cite{ericson}, where certain fluctuations and correlations
of cross sections are predicted. An open system can be studied by
the effective non-Hermitian Hamiltonian \cite{feshbach,MW69} that describes
the intrinsic dynamics coupled to the continuum. It turns out \cite{Zel1}
that transition from isolated to overlapping resonances implies
the collectivisation of overlapped states interacting through continuum.
For a small number $M$ of open channels, the restructuring of the widths
leads to the segregation of $M$ short-lived states while the remaining
states acquire narrow widths and long lifetime. This transition is similar
to the optical super-radiance \cite{dicke54}.

One of the brightest examples of quantum phenomena in open mesoscopic systems
is given by the universal conductance fluctuations \cite{been,imry99}.
There exist well pronounced similarities and some differences between them
and nuclear Ericson fluctuations \cite{weiden90}. Some aspects of this
interrelation were studied in our previous work \cite{CIZB07}. Using the model
of interacting fermions analogous to the nuclear continuum shell model
\cite{CSM06}, we analysed the behaviour of the system in function of
the intrinsic interaction strength, coupling to the continuum, and number
of open channels. Below we study in detail how many-body chaotic dynamics
inside the system is translated into observable features of many-channel
signal transmission.
%%%%%%%%%%%%%%%%%%%%%%%%%%%%%%%%%%%%%%%%%%%%%%%%%%%%%%%%%%%%%%%%%%%%%%%%%
\section{Intrinsic chaos}
Our model describes $n$ interacting fermions that occupy $m$ single-particle
levels of energies $\epsilon_s$. The intrinsic many-body Hamiltonian can be
written as
\begin{equation}
\label{Hmb}H_{\lambda} = H_0 + \widetilde{\lambda} V,
\end{equation}
where $H_0$ stands for the mean-field part describing non-interacting
particles (or quasi-particles), and $V$ contains the two-body interaction
between the particles with the variable strength $\tilde{\lambda}$.
The matrix $H_{\lambda}$ of size $N=m!/[n!(m-n)!]$ is constructed in
the many-particle basis $|k\rangle$ of the Slater determinants,
$\left| k\right\rangle =a_{s_1}^{\dagger}\,.\,\,.\,\,.\,a_{s_{n}}^{\dagger }
\left| 0\right\rangle$, where $a_{s}^{\dagger }$ and  $ a_{s}$ are the
creation and annihilation operators,
\begin{equation}
\label{H0V}H_0 =
\sum_{s=1}^{m} \epsilon_{s}\,a_{s}^{\dagger }a_{s};
\:\:\:\:\:\:\:\:\:\:\:\:\:\:\:\:\:\:\: V =\frac 12\sum \widetilde{V}_{s_1 s_2
s_3 s_4}\,a_{s_1}^{\dagger}a_{s_2}^{\dagger} a_{s_3}a_{s_4} .
\end{equation}
Each many-body matrix element $V_{lk}=\left<l|V|k\right>$ is a sum of a number
of two-body matrix elements $\widetilde{V}_{s_1 s_2 s_3 s_4}$ involving at
most four single-particle states $|s\rangle$. For this reason, many matrix
elements $V_{lk}$ vanish, and the matrix $V$ is very different from random
matrices of the Gaussian orthogonal ensemble (GOE); for details, see, for
example, \cite{brody81,FI97}. The ordered single-particle energies,
$\epsilon_s$, are assumed to have a Poissonian distribution of spacings, with
the mean level density $1/d_{\lambda}$,  implying regular dynamics of the
non-interacting system. The interaction $V$ belongs to an ensemble that is
characterised by the variance of the normally distributed two-body random
matrix elements, $\langle \widetilde{V}_{s_1 s_2 s_3 s_4}^2\rangle$, and
normalised in such a way that $\langle V_{l,k}^2\rangle=1$.

It is known \cite{A90,AGKL97,FI97} that chaotic properties of
the {\it two-body random interaction} (TBRI) Hamiltonians of
the type (\ref{H0V}) are determined by  the control parameter
$\lambda = \widetilde{\lambda} /d_f$ where
$d_f$ is the mean energy spacing between many-body states
directly coupled by the two-body interaction. Note that $d_f
\gg D_0$ where $D_0$ is the mean level spacing between
many-body states. The properties of the spectra, eigenstates
and observables in the model (\ref{Hmb}) have been
thoroughly studied in \cite{FI97}. In particular, it was shown
that the critical value for the onset of strong {\it many-body
chaos} is determined by the condition $ \lambda >
\lambda_{cr} \approx 2(m-n)/N_f $ where $N_f=n(m-n)+n(n-1)(m-n)(m-n-1)/4$.

To compare with the above model of many-body chaos, we also
consider the standard random matrix model typically used to
describe the onset of {\it one-body chaos}.
The corresponding Hamiltonian has the form,
\begin{equation}
\label{Hob}H_{\mu} = H^{\circ} + \widetilde{\mu} H_{GOE}.
\end{equation}
Here $H^{\circ}$ is a diagonal matrix with the Poissonian
distribution of spacings between its ordered eigenvalues, and
$H_{GOE}$ is a $N\times N$ matrix belonging to the GOE.
In such a description, the Hamiltonian $H_{\mu}$ can be
treated as a generic model describing an electron in a quantum
dot, or as a model of optical or electromagnetic waves in a
closed cavity with bulk disorder. The control
parameter, $\mu=\widetilde{\mu}/d_{\mu} \sqrt{N}$, determining
degree of one-body chaos is the ratio of the variance of
matrix elements of $H_{GOE}$ to
the mean energy level spacing $d_{\mu}$ between the eigenstates of
$H^{\circ}$. The transition to strong chaos occurs for $\mu >
\mu_{cr} \approx 1$.

%%%%%%%%%%%%%%%%%%%%%%%%%%%%%%%%%%%%%%%%%%%%%%%%%%%%%%%%%%
\section{Coupling to continuum}
Our aim is to study the statistical properties of {\sl open}
systems with internal dynamics described by above two Hamiltonians.
According to the well developed formalism \cite{MW69,AWM75,VWZ85},
scattering properties of an open system can be formulated with the
effective non-Hermitian Hamiltonian ${\cal H}$,
\begin{equation}
{\cal H}= H - \frac{i}{2}\, W\,;\,\,\,\,\,\,\,\,\,\,\, W_{ij}=\sum
_{c=1}^M A_{i}^{c}A_j^c,                         \label{Hnon}
\end{equation}
where $H$ is either $H_{\lambda}$ or $H_{\mu}$.
Here we neglect an additional Hermitian term (the principal value of
the dispersion integral) that appears in the elimination of the continuum
\cite{R91,CSM06}. We consider the middle of the energy spectrum, where
this term vanishes. The non-Hermitian part, $W$, describes the coupling
between $N$ intrinsic states $|i\rangle,|j\rangle,$ through $M$ open decay
channels labelled as $a,b,c...$. The factorised structure of $W$
is dictated by the unitarity of the scattering matrix. We restrict ourselves
by time-invariant systems, thus the transition amplitudes $A_{i}^{c}$
between intrinsic states $|i\rangle$ and channels $c$ are real.

The amplitudes $A_i^c$ are assumed to be random independent Gaussian
variables with zero mean and variance
\begin{equation}
\langle A_i^c A^{c'}_j\rangle=\delta_{ij}\delta_{cc'}\,\frac{
\gamma^{c}}{N}.                                \label{2}
\end{equation}
This is compatible with the GOE or TBRI models where generic intrinsic
states coupled to continuum have a very complicated structure, while
the decay probes specific simple components of these states related to
a finite number of open channels (see discussion in \cite{CIZB07}).
Below we neglect a possible energy dependence of the amplitudes that is
important near thresholds and is taken into account in realistic shell
model calculations \cite{CSM06}. The effective parameter determining
the strength of the continuum coupling can be written as
\begin{equation} \kappa^{c}=\frac{\pi\gamma^{c}}{2ND}, \label{3a}
\end{equation}
and we consider $M$ equiprobable channels, $\gamma^{c}=\gamma, \,
\kappa^{c}=\kappa$.

All scattering properties of the system with the non-Hermitian Hamiltonian
(\ref{Hnon}) are determined by the scattering matrix,
$S^{ab}=\delta_{ab}-i{\cal T}^{ab}$,
with
\begin{equation}
{\cal T}^{ab}(E)=\sum_{i,j}^N A_i^a\left(\frac{1}{E-{\cal H}}
\right)_{ij} A_j^b.                       \label{5}
\end{equation}
The complex eigenvalues ${\cal E}$ of ${\cal H}$ coincide with the poles
of the $S$-matrix and, for small $\gamma$, determine energies and widths
of {\it isolated} resonances. In the critical region $\kappa \approx 1$ with
crossover to {\sl overlapping} resonances, the width distribution displays
sharp segregation of broad short-lived ({\sl superradiant}) states and
very narrow long-lived ({\sl trapped}) states \cite{Zel1,ISS94}.
Correspondingly, the distribution of poles of the scattering matrix undergoes
a transition from one to two ``clouds" of poles in the complex plane of
resonance energies \cite{haake}; the number of broad states coincides
with the number $M$ of open channels (the rank of the matrix $W$).
For the model (\ref{Hmb}), the statistical properties of resonances as
a function of the interaction between particles and the coupling to continuum
have been thoroughly studied in earlier papers \cite{CIZB07,CISZB08}.

Our main interest is in the dependence of fluctuation and correlation
properties of scattering on the degree of internal chaos and strength of
continuum coupling. The average values of reaction cross sections,
$\sigma^{ab}(E)$, are fully defined by the transition amplitudes
${\cal T}^{ab}(E)$,
\begin{equation}
\sigma^{ab}(E) = |{\cal T}^{ab}(E)|^2.            \label{4}
\end{equation}
In our notations the cross sections are dimensionless since we omit the
common factor $\pi/k^{2}$. We will use the terminology borrowed from
nuclear physics referring to the $b=a$ process as ``elastic'' and $b\neq a$
as ``inelastic'' although all reactions are considered within the fixed energy
window; in what follows we study both types of cross sections.
We ignore the smooth potential phases irrelevant for our purposes
leaving only properties due to the compound resonances and therefore to
the intrinsic dynamics. According to the theory of Ericson fluctuations
\cite{ericson}, the scattering amplitude of any process can be written
as the sum of the average and fluctuating parts, ${\cal T}^{ab}(E)=
\langle{\cal T}^{ab}(E)\rangle + {\cal T}^{ab}_{{\rm fl}}(E)$, with
$\langle{\cal T}^{ab}_{{\rm fl}}(E)\rangle =0$. The average cross section,
$\sigma= |{\cal T}|^2$, can be also divided into two contributions,
$\langle\sigma\rangle=\langle\sigma_{{\rm dir}}\rangle + \langle
\sigma_{{\rm fl}}\rangle$. Here the direct reaction cross section,
$\langle \sigma_{{\rm dir}} \rangle$, is determined by the average scattering
amplitude only, while $\langle \sigma_{{\rm fl}} \rangle$ is the fluctuational
part also known as the compound nucleus cross section.

%%%%%%%%%%%%%%%%%%%%%%%%%%%%%%%%%%%%%%%%%%%%%%%%%%%%%%%%%%%
\section{Cross section correlations}
The fluctuations of both elastic and inelastic cross sections strongly depend
on the coupling to the continuum. According to the standard Ericson theory,
in the region of strongly overlapping resonances, $\kappa \approx 1$,
the variance of both elastic and inelastic cross sections for large $M\gg 1$
can be expressed via the average cross sections, $ {\rm Var}(\sigma)=
\langle\sigma_{{\rm fl}}\rangle ^2$. Our data for the many-body Hamiltonian
(\ref{Hmb}) confirm this expectation for $M\geq 10$ for inelastic cross
sections and any strength of interaction between particles. As for the elastic
cross sections, a slight dependence on an internal chaos has been found and
explained in \cite{CIZB07}.

Of special interest is the problem of correlations between
different cross sections. The commonly used quantity that is
discussed in nuclear and solid state physics, is the {\it
covariance} $C_{{\rm fl}}$ of fluctuational cross sections,
\begin{equation}
C_{{\rm fl}}= \langle \sigma_{{\rm fl}}^{ab}\sigma_{{\rm fl}}^{a'b'}\rangle  -
\langle\sigma_{{\rm fl}}^{ab}\rangle\langle \sigma_{{\rm fl}}^{a'b'}\rangle.
\label{cf_fl}
\end{equation}
In our case without direct processes we have
$\langle\sigma\rangle=\langle\sigma_{{\rm fl}}\rangle$, therefore, below we
omit the subscript ``${\rm  fl}$". The analysis of the
covariance $C$ shows that its value strongly depends on the type
of correlations. Specifically, there are 5 types of
correlations: $EE$ - elastic-elastic, when both cross sections
are elastic, $EI_1$ and $EI_0$ - elastic-inelastic correlations with
one and no common channels in the scattering, and $II_1$
and $II_0$  - inelastic-inelastic correlations with one and no
common channels, respectively.

One should stress that the theoretical analysis of the covariance
(\ref{cf_fl}) encounters serious problems even for the GOE case in place of
$H$ in Eq.~(\ref{Hnon}) (see also Ref.\cite{MS91} for different model).
The second term in Eq.~(\ref{cf_fl}) is defined by the second moments of
the scattering matrix. The corresponding expressions were obtained in
Ref. \cite{VWZ85} with the use of the super-symmetry method. In the limit
$M \gg 1$, they are reduced to the Hauser-Feshbach formula, see in
Ref. \cite{brody81}. It is much more difficult to evaluate the first term
that is determined by the four-point correlation function for matrix elements.
The only analytical expressions for this term can be found in Ref. \cite{DB2}.
However, the result obtained there is inconsistent with our numerical data,
as well as with the analysis of the universal conductance fluctuations
performed in Ref. \cite{GSNS02}.

In order to understand how the cross section correlations depend on
the strength of coupling to continuum and degree of internal chaos, we
performed a detailed numerical study of the correlations (\ref{cf_fl}) for
two models (\ref{Hmb}) and (\ref{Hob}). All data are obtained with averaging
over energy $E$ at the band centre, $-0.4 < E < 0.4$, and over a large number
of different cross sections belonging to one of the five groups defined above.

In Fig.~1 we show the correlations ($EE$) between two different elastic cross
sections, $\sigma^{aa}$ and $\sigma^{bb}$ with $a \neq b$, as a function of
the coupling parameter. We consider here two limiting cases, $\lambda,
\mu=0.2; 2.8$, for weak and strong internal chaos, respectively.
A noticeable dependence on the degree of chaos is clearly seen for both
models. There is an excellent agreement between $\lambda=2.8$ and $\mu=2.8$
for all values of $\kappa$. However, for weak coupling there is a small
difference between the $\lambda$- and $\mu$-models. Still, the general
trend of the $EE$ correlations is the same for both cases. It is interesting
that the symmetry between weak and strong coupling that is known for the GOE
models, and clearly seen here for strong chaos, is destroyed for weak chaos,
the effect lacks an analytical explanation.
\begin{figure}[h!]
%\vspace{-0.2cm}
\begin{center}
\includegraphics[width=0.38\linewidth,angle=90,bb=503 649 144 100,clip=]
{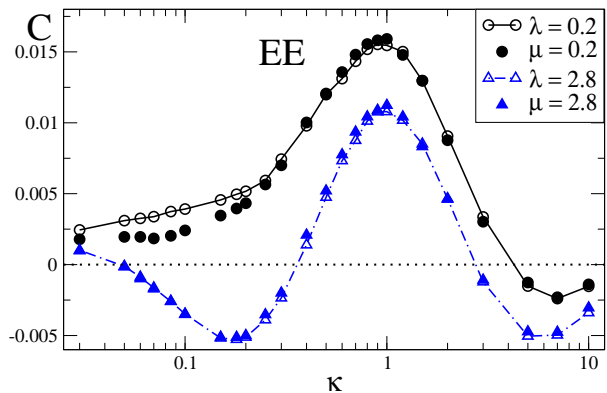}
\caption{(colour online)Dependence of $EE$ correlations on the coupling
strength, $\kappa$. Two limiting cases are shown for weak (circles) and
strong (triangles) chaos. Lines and open symbols refer to the many-body
model and full symbols to the one-body model, for $N=924$ and $M=4$ channels.
The average was done over 1000 realisations of random matrices and the error
bars (not shown) are of order of the symbol size.}
\end{center}
\label{corEE}
\vspace{-0.5cm}
\end{figure}

Next, we consider the correlations $EI_0$ between different elastic-inelastic
%and inelastic-inelastic
fluctuating cross sections with no common channel index, and the $II_0$
correlations between two different inelastic-inelastic cross sections.
These are shown in Fig.~2 for the same limiting  cases as above,
$\lambda,\mu=0.2; 2.8$. The $EI_0$ and $II_0$ correlations have a difference
by a factor close to $2$ in amplitude but follow a similar trend for all
three cases of internal chaos. Comparing the results for the two models
we arrive at a similar result as before: an excellent agreement between
$\lambda=2.8$ and $\mu=2.8$, and small deviations between the cases
$\lambda \leq 1$ and $\mu \leq 1$ for weak coupling. This is more evident
for the $EI_0$ than for the $II_0$ correlations.
\begin{figure}[h!]
%\vspace{-0.2cm}
\begin{center}
\includegraphics[width=0.44\linewidth, angle=90,bb=508 718 134 27,clip=]{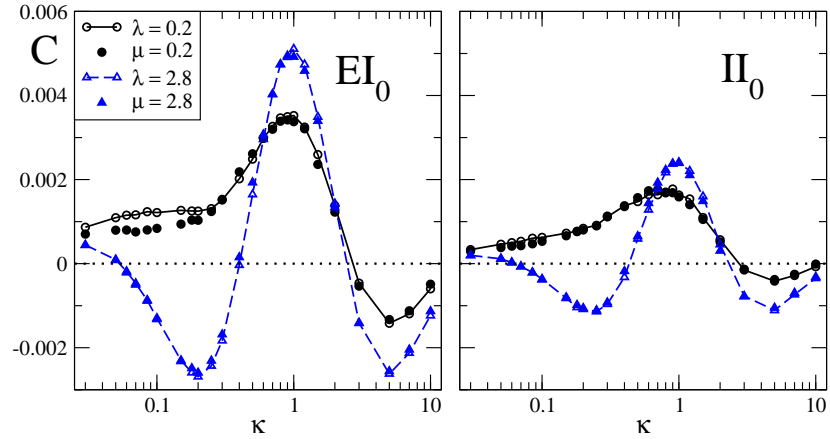}
\caption{(colour online) Dependence of the $EI_0$ and $II_0$ correlations on $\kappa$ for the parameters of Fig. 1.}
\end{center}
\label{cor0}
\vspace{-0.5cm}
\end{figure}
%\vspace{-5.cm}
The analogous case, when there is one common channel index between
the fluctuating cross sections involved in the correlation function,
is shown in Fig.~3. Here the $EI_1$ and $II_1$ correlations differ by
a factor close to $5$ in amplitude but behave similarly for both limiting
cases of internal chaos, $\lambda,\mu=0.2$ and $2.8$. Furthermore, $EI_1$
correlations are independent of the degree of internal chaos whereas
$II_1$ correlations become slightly smaller at stronger chaos. For the
$EI_1$ correlations the correspondence between the two models is
excellent for all values of $\kappa$.
\begin{figure}[h!]
\begin{center}
\includegraphics[width=0.42\linewidth,angle=90,bb=501 708 145 52,clip=]
{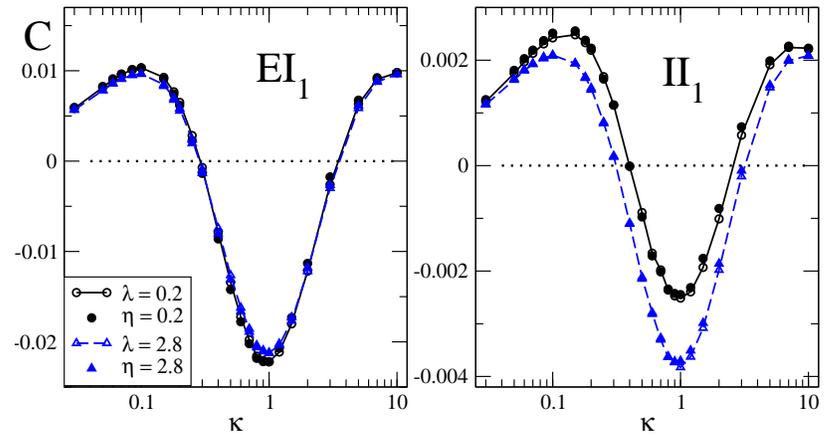}
\caption{(colour online) Dependence of the $EI_1$ and $II_1$ correlations
on $\kappa$ for the parameters of Fig. 1.}
\end{center}
\label{cor1}
\vspace{-0.5cm}
\end{figure}
We have to stress that the correlations for strong chaos with $\lambda,
\mu =2.8$, are in a good agreement
%with those
if the Hermitian part of the Hamiltonian (\ref{Hnon}) is taken from the GOE.

From our data one can see that the strongest correlations occur at {\sl perfect
coupling} to continuum, $\kappa \approx 1$. Another remarkable property is
that the correlations are either negative or positive depending on whether
there is a common channel in the correlating cross sections or such channels
are absent. Indeed, both $EI_1$ and $II_1$ correlations are negative around
$\kappa =1$, whereas the $EI_0$ and $II_0$ correlations are positive.

For a large number of channels, $M \gg 1$, these correlations are very weak,
and they are ignored in the standard Ericson theory. However, they turn out
to be very important when considering the properties of conductance
fluctuations, see below. Also, in nuclear physics there are situations when
the number of open channels is relatively small, and one can expect
that the effect of different signs of correlations can be observed
experimentally. One of the new applications of such experiments can be
the calibration of internal chaos with the use of scattering data.

Therefore, it is important to know the dependence of the correlations
(\ref{cf_fl}) on the number of channels. Our analysis has revealed that
the value of the covariance $C$ is inversely proportional to the number
${\cal N}$ of terms in each group specified by the type of correlations,
over which the averaging is performed in Eq.~(\ref{cf_fl}), see Table.
Our extensive numerical data has confirmed this dependence, $C=X/{\cal N}$
with some constants $X$ that we extracted by fitting the data with the above
dependence. This rule works perfectly starting from $M=2$ or $3$.

\vspace{0.2cm}
\begin{tabular}{|c|c|}
\hline
\multicolumn{2}{|c|}{Number of Terms, ${\cal N}$} \\
\hline
Elastic-Elastic ($EE$) & $M^2-M$\\ \hline
Elastic-Inelastic ($EI_0$) & $2(M^3-3M^2+2M)$
\\ \hline
Elastic-Inelastic ($EI_1$) & $4(M^2-M)$
\\ \hline
Inelastic-Inelastic ($II_0$) & $M^4-6M^3+11M^2-6M$
\\ \hline
Inelastic-Inelastic ($II_1$) & $4(M^3-3M^2+2M)$
\\ \hline
\end{tabular}
\vspace{0.1cm}
%%%%%%%%%%%%%%%%%%%%%%%%%%%%%%%%%%%%%%%%%%%%%%%%%%%%%%%%%%%%%%%
\section{Conductance fluctuations}
One of the most intriguing effects of mesoscopic physics is the universality of
conductance fluctuations. In order to study this effect in the framework of
our models (\ref{Hmb}) and (\ref{Hob}), one should treat $M/2$ $b$-channels
as {\it left channels} corresponding to incoming electron waves, and other
$M/2$ $a$-channels, as {\it right channels} for outgoing waves. Then, we
define the Landauer conductance $G$ in the standard way \cite{been}
omitting the common factor of $2e^{2}/h$),
\begin{equation}
G=\sum_{b=1}^{M/2} \sum_{a>M/2}^M \sigma^{ab} \label{G}.
\label{land}
\end{equation}

The properties of the conductance are entirely determined by
the inelastic cross-sections, $b\neq a$, and for equivalent
channels the average conductance (\ref{land}) reads,
\begin{equation}\label{G-av}
\langle G \rangle =\frac{M^2}{4}\langle\sigma^{ab}\rangle=
\frac{M^2}{4}\frac{T}{F+M-1}\rightarrow \frac{MT}{4}.
\end{equation}
Here $T=4\kappa/(1+\kappa)^2$ is the transmission coefficient, and
$F=\langle \sigma_{{\rm fl}}^{aa}\rangle /\langle \sigma_{{\rm fl}}^{ab}
\rangle$ is the {\sl elastic enhancement factor} \cite{harney86}. Our data
show that the value of $F$ changes from $F=2$ to $F \approx 3$ when
decreasing the strength of chaos from $\lambda, \mu = \infty$ to $\lambda,
\mu = 0$. The last expression in Eq.~(\ref{G-av}) is written for $M \gg 1$.
As one can see, the influence of internal chaos is due to the enhancement
factor $F$ only. Since there is no theory relating the enhancement factor
to the degree of chaos, we used this factor as the fitting parameter.
Our data manifest an excellent agreement with the expression (\ref{G-av})
for various values of parameters $\lambda$ and $\mu$, as well as $M$.
Note that for a very large number of channels the influence of internal
dynamics on fluctuations disappears.

As for the variance ${\rm Var}(G)$, the analytical results are available
for the GOE case only, corresponding to very large values of $\lambda$ and
$\mu$. According to different approaches (see, for example, in Ref.
\cite{guhr98}), for perfect coupling, $\kappa=1$, and very large number
of channels, $M\gg 1$, this variance takes the famous values $2/15$ and
$1/8$ for diffusive and ballistic transport, respectively. This result is
commonly considered as a striking effect of universal conductance
fluctuations. Our data reported in Fig. 4 clearly manifest that, in both
models (\ref{Hmb}) and (\ref{Hob}), for $\kappa=1$ and strong internal
chaos the value of ${\rm Var}(G)$ is close to $1/8$. A small, however,
clear difference from $1/8$ is explained by the correction due to a finite
value $M=20$. A more general expression for a finite number of channels
(for $\kappa =1$ and the GOE) can be found in Refs. \cite{MK04,been},
\begin{equation}
{\rm Var}(G)=2\frac{(M/2)^2[(M/2)+1]^2}{M(M+3)(M+1)^2},
\label{vgM}
\end{equation}
and our data perfectly agree with this result.

\begin{figure}[h!]
\begin{center}
\includegraphics[width=0.43\linewidth,angle=90,bb=497 680 140 19,clip=]
{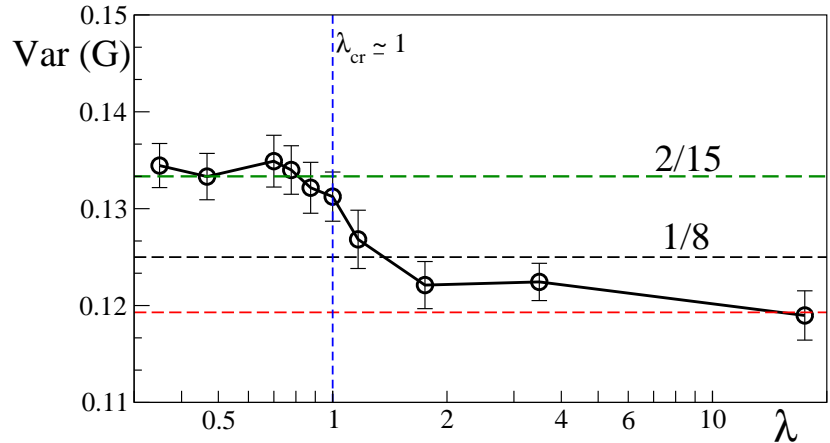}
\caption{(colour online) ${\rm Var}(G)$ versus $\lambda$ for $n=7$, $m=14$,
$N=3432$ and
$M=20$ channels for the many-body model (\ref{Hmb}).}
\end{center}
\label{fig3}
\vspace{-0.2cm}
\end{figure}
One can see from Fig.~4 that the variance ${\rm Var}(G)$
increases when $\lambda$ and $\mu$ decrease, and
crosses the ballistic value $2/15$ close to the
critical value at which the transition from weak to strong
chaos occurs in closed models.

One of the most interesting and new results of our study is how
the internal chaos influences the variance of the conductance,
${\rm Var}(G)$. As one can see from Fig.~5, in the region with
small or moderate continuum coupling, $\kappa \approx
0.1-0.5$, the value of ${\rm Var}(G)$ strongly depends on the
strength of inter-particle interaction, $\lambda$, in the model
(\ref{Hmb}) of many-body chaos, and on the perturbation
parameter $\mu$ in the model (\ref{Hob}) of one-body chaos.
The strongest influence of chaos occurs for $\kappa \approx 0.2$,
and the results are practically the same for both models, provided
the appropriate normalisation of $\lambda$ and $\mu$ is made.
This important result may find various applications in theory of
conductance fluctuations. In particular, one may try to extract
information about internal dynamics from experimental data when
changing the degree of coupling to continuum (for example, the degree
of openness of quantum dots).

\begin{figure}[h!]
%\vspace{-0.5cm}
\begin{center}
\includegraphics[width=0.37\linewidth,angle=90,bb=465 768 170 1,clip=]
{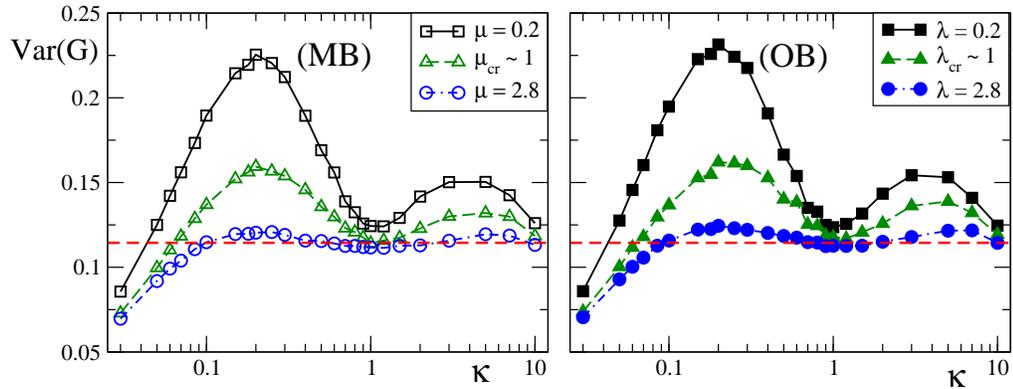}
\caption{(colour online) Variance of the conductance for the many-body model
(\ref{Hob}),
left panel, as compared to
the one-body model, right panel. The average was done over 700 realisations
of random matrices with $N=924$ and $M=10$.}
\end{center}
\label{fig4}
\end{figure}
An important feature of the data in Fig.~5 is that for $\kappa \approx 1$
the influence of intrinsic chaos is weak. This can be explained as follows.
When coupling is perfect, an
%direct
interaction with the continuum through forming broad states in
various channels is very strong in comparison with an internal process of
chaotization, therefore, the latter may be neglected. We found that for
$\kappa \approx 1$ the sensitivity of the conductance fluctuations to
the degree of chaos decreases with an increase of number of channels $M$.

It is instructive to show that the main properties of conductance fluctuations
cannot be explained if one neglects cross section correlations discussed
above. For the first time, the role of these correlations in application
to conductance fluctuations has been discussed in Refs.
\cite{LSF87,FKLS88,MAS88}. The variance ${\rm Var}(G)$ can be rewritten as
%follows
\begin{equation}
\label{var+corr}
{\rm Var}(G) =
\frac{M^2}{4}\left(\frac{T}{F+M-1}\right)^2 + N_1
C_1 + N_0 C_0 ,
\end{equation}
where $N_1=L(M-2), \,N_0=L(L-M+1),\, L=M^2/4$ and the terms $C_1$ and $C_0$
stand for the $II_1$ and $II_0$ correlation functions, respectively, see
above. If one neglects the correlations, the first term gives $1/4$
(for $T=1$ and $M\gg 1$), instead of $1/8$. Our analysis shows that for
the strong interaction and $M \gg 1$ one obtains $C_1 \approx -M^{-3}$ and
$C_0 \approx 2 M^{-4}$. Therefore, in the limit of a large number of channels,
the first term in r.h.s. of Eq. (\ref{var+corr}) that equals $1/4$, is
cancelled by the second term, and the third term tends to $1/8$ resulting in
${\rm Var} (G) =1/8$.

This result clearly demonstrates the crucial role of correlations determining
the conductance fluctuations (see, also, Ref. \cite{GSNS02}). Remarkably,
the $II_1$-correlations cancel the first term $1/4$, and the value $1/8$ is
due to the $II_0$-correlations (term $C_0$) only. A highly non-trivial role
of the $C_1$ and $C_0$ terms can be also manifested in the correlations of
speckle pictures \cite{FKLS88}.
%%%%%%%%%%%%%%%%%%%%%%%%%%%%%%%%%%%%%%%%%%%%%%%%%%%%%
\section{Conclusions}
 To conclude, we studied the interplay of complicated intrinsic dynamics
and coupling to the outside world for typical quantum systems with one-body
or many-body intrinsic chaos, the first one coming from the disordered
single-particle spectrum and the second one emerging as a result of
inter-particle interactions. The openness of the system is described by
the effective non-Hermitian Hamiltonian that fully respects the unitarity
requirements and allows to calculate, in the same framework, cross sections
of various processes, their fluctuations and correlations. As a
manifestation of the general features of underlying physics, the models are
equally valid for description of nuclear reactions with the transition from
isolated to overlapping resonances and for conductance fluctuations in
mesoscopic condensed matter devices. We found that the correlations
of inelastic cross sections are very different for the processes with and
without a common channel, being negative in the latter case in the region
of perfect coupling when the typical decay widths and resonance spacings are
of the same magnitude. Another important result is that for the
conductance fluctuations the dependence on the degree of intrinsic chaos is
strong at intermediate continuum coupling, in contrast with the region of
perfect coupling, for which the continuum dominates, and the
fluctuations are known to be independent of internal dynamics. Many results
of the conductance theory are numerically confirmed being explained by
the specific correlations of partial cross sections of very general
origin.
\vspace{-0.4cm}
\acknowledgments
\vspace{-0.4cm}
We gratefully acknowledge stimulating discussions with Y.Alhassid,
B.L.Altshuler, P.Mello, A.Richter and H.Weidenm\"uller. F.M.I. and V.Z. thank the INT at University of Washington for hospitality and support; V.G.Z. and S.S.
acknowledge support from the NSF grant PHY-0758099 and The Leverhulme Trust,
respectively. The work of G.P.B. was  carried out under the auspices of
the National Nuclear Security Administration of the U.S. Department of Energy
at Los Alamos National Laboratory under Contract No. DEAC52-06NA25396.
The work of F.M.I. was partly supported by CONACyT grant No. 80715.


\begin{thebibliography}{0}

%\bibitem{b.a}
%  \Name{Author F., Author S. \and Author T.}
%  \REVIEW{Some Rev. A}{69}{1969}{9691}.

%\bibitem{b.b}
%  \Name{Author F. \and Author S.}
%  \Book{Some Book of Interest}
%  \Editor{A. Editor}
%  \Vol{9}
%  \Publ{Publishing house, City}
%  \Year{1939}
%  \Page{666}.

%\bibitem{b.c}
%  \Editor{Editor A.}
%  \Book{Some Book of Interest}
%  \Vol{9}
%  \Publ{Publishing house, City}
%  \Year{1939}
%  \Section{A}.

\bibitem{MK04}
\Name{P.A. Mello \and N. Kumar}
\Book{Quantum Transport in Mesoscopic Systems: Complexity and
  Statistical Fluctuations}
\Publ{Oxford University Press, Oxford}
\Year{2004}.

\bibitem{brody81}
\Name{ T.A. Brody, J. Flores, J.B. French, P.A. Mello, A. Pandey \and
S.S.M. Wong}
\REVIEW{Rev. Mod. Phys.}{53}{1981}{385}.

\bibitem{grib94}
\Name{ V.V. Flambaum, A.A. Gribakina, G.F. Gribakin, and M.G. Kozlov}
\REVIEW{Phys. Rev. A}{50}{1994}{267}.

\bibitem{big}
\Name{V. Zelevinsky, B.A. Brown, N. Frazier, and M. Horoi}
\REVIEW{Phys. Rep}{276}{1996}{85}.

\bibitem{guhr98}
\Name{T. Guhr, A. M\"{u}ller-Groeling, and H.A.Weidenm\"{u}ller}
\REVIEW{Phys. Rep.}{299}{1998}{189}.

\bibitem{PW07}
\Name{T. Papenbrock and H.A. Weidenm\"{u}ller}
\REVIEW{ Rev. Mod. Phys.} {79}{2007}{997}.

\bibitem{FI97}
\Name{V.V. Flambaum and F.M. Izrailev}
\REVIEW{Phys. Rev. E}{56}{1997}{5144}.

\bibitem{A90}
\Name{S. Aberg}
\REVIEW{Phys. Rev. Lett.}{26}{1990}{3119}.

\bibitem{AGKL97}
\Name{B.L. Altshuler, Y. Gefen, A. Kamenev and L.S. Levitov}
\REVIEW{Phys. Rev. Lett.}{78}{1997}{2803}.

\bibitem{ericson}
\Name{T. Ericson}
\REVIEW{Ann. Phys.}{23}{1963}{390};
\Name{D. Brink and R. Stephen}
\REVIEW{Phys. Lett.}{5}{1963}{77};
\Name{T. Ericson and T. Mayer-Kuckuk}
\REVIEW{Ann. Rev. Nucl. Sci.}{16}{1966}{183}.

\bibitem{feshbach}
\Name{H. Feshbach}
\REVIEW{Ann. Phys.}{5}{1958}{357};
\REVIEW{Ann. Phys.}{19}{1962}{287};
\REVIEW{Rev. Mod. Phys.} {36}{1964}{1076}.

\bibitem{MW69}
\Name{C. Mahaux and H.A. Weidenm\"{u}ller}
\Book{Shell Model Approach to Nuclear Reactions}
\Publ{North Holland, Amsterdam}
\Year{1969}.

\bibitem{Zel1}
\Name{V. V. Sokolov and V. G. Zelevinsky}
\REVIEW{Phys. Lett. B} {202}{1998}{10};
\REVIEW{Nucl. Phys.}{A504}{1989}{562} (1989);
\REVIEW{Ann. Phys. (N.Y.)}{216}{1992}{323}.

\bibitem{dicke54}
\Name{R.H. Dicke}
\REVIEW{Phys. Rev.}{93}{1954}{99}.

\bibitem{been}
\Name{C.W.J. Beenakker}
\REVIEW{Rev. Mod. Phys.}{69}{1997}{731}.

\bibitem{imry99}
\Name{Y. Imry \and R. Landauer}
\REVIEW{Rev. Mod. Phys.}{71}{1999}{S306}.

\bibitem{weiden90}
\Name{H.A. Weidenm\"{u}ller}
\REVIEW{Nucl. Phys.}{A518}{1990}{1}.

\bibitem{CIZB07}
\Name{G.L. Celardo, F.M. Izrailev, V.G. Zelevinsky \and G.P. Berman}
\REVIEW{Phys. Rev. E}{76}{2007}{031119};
\REVIEW{Phys. Lett. B}{659}{2008}{170}.

\bibitem{CSM06}
\Name{A. Volya \and V. Zelevinsky}
\REVIEW{Phys. Rev. C}{74}{2006}{064314};
\REVIEW{Phys. Rev. Lett.}{94}{2005}{052501}.

\bibitem{AWM75}
\Name{D. Agassi, H.A. Weidenm\"{u}ller, \and G. Mantzouranis}
\REVIEW{Phys. Rep.}{3}{1975}{145}.

\bibitem{VWZ85}
\Name{J.J.M. Verbaarschot, H.A. Weidenm\"{u}ller, \and M.R. Zirnbauer}
\REVIEW{Phys. Rep.}{129}{1985}{367}.

\bibitem{R91}
\Name{I. Rotter}
\REVIEW{Rep. Prog. Phys.}{54}{1991}{635}.

\bibitem{ISS94}
\Name{F.M. Izrailev, D. Sacher \and V.V. Sokolov}
\REVIEW{Phys. Rev. E}{49}{1994}{130}.

\bibitem{haake}
\Name{F. Haake {\sl et al.}}
\REVIEW{Z. Phys. B }{88}{1992}{359}.

\bibitem{CISZB08}
\Name{G.L. Celardo, F.M. Izrailev, S. Sorathia, V.G. Zelevinsky, \and
G.P. Berman}
\Book{Nuclei and Mesoscopic Physics 2007}
\Editor{P. Danielewicz, P. Piecuch, \and V. Zelevinsky},
\REVIEW{AIP Conference Proceedings}{995}{2008}{75}.

\bibitem{MS91}
\Name{P.A.Mello \and A.D.Stone}
\REVIEW{Phys. Rev. B}{44}{1991}{3559}.

\bibitem{DB2}
\Name{E.D. Davis \and D. Boose}
\REVIEW{Z. Phys.}{A332}{1989}{427}.

\bibitem{GSNS02}
\Name{A. Garc\'ia-Mart\'in, F. Scheffold, M. Nieto-Vesperinas, \and
J.J. S\'aenz}
\REVIEW{Phys. Rev. Lett.}{88}{2002}{143901}.

\bibitem{harney86}
\Name{H.L. Harney, A. Richter, \and H.A. Weidenm\"{u}ller}
\REVIEW{Rev. Mod. Phys.}{58}{1986}{607}.

\bibitem{LSF87}
\Name{P.A.Lee, A.S.Stone, \and H.Fukuyama}
\REVIEW{Phys. Rev. B}{35}{1987}{1039}.

\bibitem{FKLS88}
\Name{S.Feng, C.Kane, P.A.Lee, \and A.D.Stone}
\REVIEW{Phys. Rev. Lett.}{61}{1988}{834};
\Name{I.Freud, M.Rosenbluth, \and S.Feng}
\REVIEW{ibid}{61}{1988}{2328}.

\bibitem{MAS88}
\Name{P.A.Mello, E.Akkermans, \and B.Shapiro}
\REVIEW{Phys. Rev. Lett.}{61}{1988}{459}.

\end{thebibliography}
\end{document}